%% file: nn_re_distribs_2n_halos.tex
\documentclass[prc,superscriptaddress,nofootinbib,noeprint,dvipsnames,twocolumn]{revtex4-2}

\usepackage[english]{babel}
\usepackage[hidelinks]{hyperref}
\makeatletter\AtBeginDocument{\let\@elt\relax}\makeatother

\usepackage{booktabs}
\usepackage{longtable}
\usepackage{tabularx}
\usepackage{graphicx}
\usepackage{amsmath}
\usepackage{float}
\usepackage[english,capitalise]{cleveref}
\usepackage{mathtools}
\usepackage{bbm}

\usepackage[utf8]{inputenc}

\usepackage{siunitx}
\usepackage[dvipsnames,usenames]{xcolor}
\definecolor{green}{rgb}{0.2,0.6,0.2}

\usepackage{physics}
\usepackage{csquotes}
\usepackage{multirow}
\usepackage{ifthen}
\usepackage{ulem}

\makeindex

\usepackage{our_newcommands0722}

\newcommand{\orcid}[1]{\href{https://orcid.org/#1}{\includegraphics[scale=0.055]{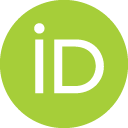}}}

\renewcommand{\vec}[1]{\v{#1}}

\begin{document}

\title{\texorpdfstring{Universality of \(nn\) distributions of \(s\)-wave \(2n\) halos and the unitary limit}{Universality of nn distributions of s-wave 2n halos and the unitary limit}}

\input{tex/authors.tex}


\date{September 7, 2023}

\begin{abstract}
We calculate neutron-neutron relative-energy distributions of \(s\)-wave two-neutron (\(2n\)) halo nuclei 
using Halo Effective Field Theory (Halo EFT) at leading order.
At this order these systems are described by the \(2n\) separation energy, the neutron-core (\(nc\)) virtual-state energy and the neutron-neutron (\(nn\)) scattering length.
We focus on knockout reactions where the removal of the core is sudden, such that the final-state interactions are dominated by the \(nn\) interaction.
We consider the neutron relative-energy distribution for the nuclei \(^{11}\)Li, \(^{14}\)Be, \(^{17}\)B, \(^{19}\)B, and \(^{22}\)C.
We show that the ground-state neutron momentum distributions of all these nuclei stem from a single curve, which can be obtained by taking both the neutron-core and neutron-neutron interaction to the unitary limit. 
This universal description can be extended to the final distribution measured in experiment by including \(nn\) final-state interactions via the approximate technique of enhancement factors.
For all the nuclei considered we find good agreement between the full leading-order Halo EFT calculation and the universal prediction obtained in this way.
The universality of the ground-state momentum distribution in two-neutron Borromean halos can thus be tested by dividing the experimental results from sudden core knockout by the enhancement factor and comparing to the unitary-limit prediction.
\end{abstract}
\maketitle

\input{tex/introduction.tex}

\input{tex/foundations.tex}

\input{tex/universality.tex}

\input{tex/main_part.tex}

\input{tex/conclusion.tex}

\acknowledgments

We thank T. Aumann for useful discussions.
M.G. and H.W.H. acknowledge support by Deutsche Forschungsgemeinschaft (DFG, German Research Foundation) - Project-ID 279384907 - SFB 1245.
H.W.H. has been supported by the German Federal Ministry of Education and Research (BMBF) (Grant No. 05P21RDFNB).
D.R.P. was supported by the US Department of Energy, contract DE-FG02-93ER40756.
M.G. thanks Ohio University and its Institute for Nuclear and Particle Physics, where parts of this research have been conducted, for the hospitality and supporting his visit.

\bibliographystyle{apsrev4-2}
\bibliography{he6_and_nn_scattering_length.bbl}

\end{document}

%% file: tex/authors.tex

\author{Matthias G\"{o}bel \orcid{0000-0002-7232-0033}}
\email[E-mail: ]{goebel@theorie.ikp.physik.tu-darmstadt.de}
\affiliation{Technische Universit\"{a}t Darmstadt, Department of Physics, Institut f\"{u}r Kernphysik, 64289 Darmstadt, Germany}

\author{Hans-Werner Hammer \orcid{0000-0002-2318-0644}}
\affiliation{Technische Universit\"{a}t Darmstadt, Department of Physics, Institut f\"{u}r Kernphysik, 64289 Darmstadt, Germany}
\affiliation{ExtreMe Matter Institute EMMI and Helmholtz Forschungsakademie Hessen für FAIR (HFHF), GSI Helmholtzzentrum f\"{u}r Schwerionenforschung GmbH, 64291 Darmstadt, Germany}

\author{Daniel R. Phillips \orcid{0000-0003-1596-9087}}
\affiliation{Institute of Nuclear and Particle Physics and Department of Physics and Astronomy, Ohio University, Athens, OH 45701, USA}

%% file: tex/introduction.tex

\section{Introduction}

Halo nuclei are a species of exotic nuclei living in the nuclear chart beside the valley of stability.
They are characterized by a separation between a more tightly bound cluster, called a core, and one to multiple more loosely bound nucleons, which form the halo \cite{Riisager:2012it,Hammer:2017tjm}.
Neutron as well as proton halos have been observed.
Neutron halos seem to occur more frequently, as they are not subject to the long-range Coulomb interaction.
Neutron halos were discovered in the 1980s at radioactive beam facilities by measuring the markedly larger cross section associated with their greater spatial extension~\cite{Tanihata:2016zgp}.
This, in turn,  is related to the small separation energy of the halo neutrons~\cite{Hansen:1987mc}.

This separation of scales between the binding of the halo nucleons and the energy scales of the core, such as excitation energies or nucleon removal energies, form the basis for a systematized cluster description of these systems that uses the methodology of effective field theories (EFTs).
This so-called Halo EFT was first formulated for \(^{5}\)He \cite{Bertulani:2002sz,Bedaque:2003wa} and has since then been applied to a number of nuclei and different observables.
Halo EFT can be understood as an implementation of an expansion around the unitary limit of infinite two-body scattering length \cite{Hammer:2010kp}.
An in-depth review of Halo EFT was given in Ref.~\cite{Hammer:2017tjm}.
The underlying idea is that observables can be expanded in powers of a low-momentum scale over a high-momentum scale.
This high-momentum scale is also called the breakdown scale and is set by the lowest scale of omitted physics, e.g., the excitation energy of the core.
A key ingredient is the power counting which determines the order at which different parts of the interaction contribute in this expansion.
In this way, the observables can be calculated systematically up to a certain order with a clear perspective on how this result could be improved, i.e., by going an order higher.
Moreover, the theoretical uncertainties of the results can be estimated from the anticipated size of omitted higher order terms.

Many observables of halo nuclei, structural properties as well as those related to reactions, have been calculated in Halo EFT.
In this paper, we want to look at the \(nn\) relative-energy distributions of \(s\)-wave \(2n\) halos after core knockout.
From the theoretical side this observable has already been investigated for the \(p\)-wave \(2n\) halo \(^6\)He \cite{Gobel:2021pvw}.
In that case, this was done with specific emphasis on inferring the \(nn\) scattering length from this distribution.
An experiment to extract the not-yet precisely known \(nn\) scattering length is approved at RIKEN RIBF.
Since Ref. \cite{Gobel:2021pvw} has shown that this observable can be theoretically calculated and it is also experimentally feasible, here we are interested in calculating the same observable for different \(s\)-wave \(2n\) halo nuclei.

The assumptions regarding the reaction are the same as for the reaction \(^6\)He\((p, p'\alpha)nn\) discussed in Ref. \cite{Gobel:2021pvw}.
It is assumed that the \(2n\) halo \(h \equiv [cnn]\) is shot on a hydrogen target at high velocity so that the core \(c\) is knocked out and the two neutrons continue to fly along the original path until they get detected.
This reaction is denoted as \(h(p, p'c)nn\).
Through the sudden knockout of the core, the halo neutrons are almost unperturbed by the core and \(nn\) final-state interaction (FSI) is by far the most important FSI.

In this study, we investigate the two-neutron halos \(^{11}\)Li, \(^{14}\)Be, \(^{17}\)B, \(^{19}\)B, and \(^{22}\)C.
They all display a low-lying \(nc\) \(s\)-wave virtual state, whose energy parameterizes the \(nc\) interaction at leading order.
Other leading-order interactions are the \(s\)-wave \(nn\) interaction given by the scattering length as well as a three-body force adjusted to reproduce the physical two-neutron separation energy.
The characteristic properties of the considered halos, such as the mass number \(A\) of the halo core, the two-neutron separation energy \(S_{2n}\), and the virtual-state energy  \(E_{nc}^*\) of the \(nc\) subsystem, are listed in \cref{tab:char_param}.
The numerical values for \(S_{2n}\) and \(E_{nc}^*\) are taken from the 2020 Atomic Mass Evaluation~\cite{Wang:2021xhn} except for \(S_{2n}\) of \(^{22}\)C, where we use the value 100~keV from Ref.~\cite{Hammer:2017tjm} based on a different data set instead of the 35~keV quoted in Ref. \cite{Wang:2021xhn}.
 
\begin{table}[hbt]
  \caption{Characteristic properties of the considered two-neutron halo nuclei.
    The values for \(S_{2n}\) and for the virtual-state energy \(E_{nc}^*\) are
    taken from Ref. \cite{Wang:2021xhn}.
    The only exception is \(S_{2n}\) of \(^{22}\)C, for which we use the value
    given in Ref. \cite{Hammer:2017tjm}.
  }\label{tab:char_param}
  \begin{tabular}{lccc}
    \toprule
      nucleus & \(A\) & \(S_{2n}\) [keV] & \(E_{nc}^*\) [keV]\\
    \midrule
      \(^{11}\)Li & 9  &  369 &  26 \\
      \(^{14}\)Be & 12 & 1266 & 510 \\
      \(^{17}\)B  & 15 & 1384 &  83 \\
      \(^{19}\)B  & 17 &   90 &   5 \\
      \(^{22}\)C  & 20 &  100 &  68 \\
    \bottomrule
  \end{tabular}
\end{table}

We will investigate how universal the neutron energy distributions of these different nuclei are. 
In Refs.~\cite{Hiyama:2019bzy,Hiyama:2022loc} Hiyama et al. argued that the \(^{17}\)B-\(n\)-\(n\) system is near the unitary limit.
In this study we go further, and argue that a universal prediction for all---or at least many---\(s\)-wave two-neutron halos can be obtained by taking all two-body interactions, the \(nn\) interaction as well as the \(nc\) interaction, in the unitary limit.
In our leading-order EFT description, this corresponds to setting the respective inverse scattering length to zero.
Halo EFT also includes a core-neutron-neutron interaction at leading order, and this ``three-body'' force is tuned to reproduce the two-neutron separation energy of each halo we consider. 

Our philosophy is thus similar to that of K\"onig et al. who propose to understand the properties of light nuclei in pionless EFT in a perturbative expansion around a leading order given by two-body interactions in the unitary limit plus a three-body force \cite{Konig:2016utl}.
They demonstrated that this works well for nuclei with \(A \leq 4\) and suggested examining the expansion also in higher-\(A\) systems.
The proposal of K\"onig et al. was further investigated in Refs.~\cite{Vanasse:2016umz,Contessi:2017rww,vanKolck:2017jon,Gattobigio:2019omi,Dawkins:2019vcr,Deltuva:2020aws,Contessi:2022vhn}.

Our work can be seen as an extension of the ideas of Ref.~\cite{Konig:2016utl} to clustered systems that focuses on the 
\(nn\) relative-energy distribution. 
We demonstrate that this observable can be accurately computed via a unitary-limit treatment of the halo nucleus ground state as follows.
In Sec.~\ref{ssec:recapitulate} we recapitulate how a leading-order Halo EFT calculation of a two-neutron halo is 
carried out, displaying the Faddeev equations, and explaining how to combine the particle-dimer amplitudes that are the output of those equations to obtain the full three-body wave function.
In Sec.~\ref{ssec:res_distribs} we define the neutron relative-energy distribution in terms of that wave function, discuss how \(nn\) final-state interactions are included in its calculation, and present our results for this distribution for each of the five different Borromean halos considered here.
Section~\ref{sec:universality} then defines 
what it would mean for these different results to be related to one universal curve.
In Sec.~\ref{sec:universalcurve} we first show that the rescaled ground-state momentum distributions of all five halos are, when plotted versus an appropriate variable, described to better than 25\% accuracy, by the unitary-limit result.
We explain why this is actually to be expected given the scales in the problem and the three-body dynamics that is at work.
We then show, in Sec.~\ref{sec:includingFSI}, that the final-state interactions that might obscure this universality can be accounted for---at a similar level of accuracy---using an enhancement factor.
We offer a summary and outlook in Sec.~\ref{sec:conclusion}.

%% file: tex/foundations.tex

\section{Halo EFT and Calculational Methods}
\label{sec:methods}

\textit{Agenda.} In order to obtain the \(nn\) relative-energy distribution we first calculate the wave function of the \(2n\) halo.
On this basis, we can either calculate the ground-state \(nn\) relative-energy distribution or we can calculate the final \(nn\) relative-energy distribution after the hard knock-out in the reaction \(h(p,p' c)nn\).
The latter final distribution includes the effect of \(nn\) FSI on the ground-state wave function.

Calculating the wave functions in Halo EFT works along the lines described in Ref. \cite{Canham:2008jd} and in Ref. \cite{Hammer:2017tjm}: one starts with the leading-order Lagrangian, which has as its degrees of freedom the neutrons and the core, together with the corresponding neutron-neutron and neutron-core dimers.
The dressed dimer propagators determine the neutron-neutron and neutron-core
$t$-matrices which enter in the Faddeev equations.
They can be obtained analytically from a Dyson equation, where the leading-order effective range parameters are employed as renormalization conditions.
While in the case of the \(nn\) interaction this is directly done using the \(nn\) scattering length \(a_{nn}\), in the case of the \(nc\) interaction the momentum \(\gamma_{nc}\) characterizing the low-lying virtual state is employed.

In the next step, the coupled integral equations for the particle-dimer amplitudes can be derived.
These equations are equivalent to the Faddeev equations. 
The amplitudes can be obtained from the equations via discretization and solving the resulting eigenvalue problem.
These correspond to the Faddeev amplitudes.
Finally, the wave functions can be calculated from the amplitudes.
The \(nn\) relative-momentum (or, equivalently relative-energy) distribution is then straightforwardly obtained from the wave function. 

\subsection{Formalism, Faddeev amplitudes, and Wave functions}
\label{ssec:recapitulate}

We use the Jacobi coordinates
\begin{align}
    \v{p}_i &\coloneqq \mu_{jk} \K{\frac{\v{k}_j}{m_j} - \frac{\v{k}_k}{m_k}} \,, \\
    \v{q}_i &\coloneqq \mu_{i(jk)} \K{ \frac{\v{k}_i}{m_i} - \frac{\v{k}_j + \v{k}_k}{M_{jk}} }\,,
\end{align}
where \(\mu_{ij} \coloneqq m_i m_j / \K{m_i + m_j}\) and \(\mu_{i(jk)} \coloneqq m_i M_{jk} / \K{m_i + M_{jk}}\).
\(M_{jk}\) is the total mass of particles \(j\) and \(k\).

The Faddeev equations for a two-neutron system with one \(nn\) and one \(nc\) interaction channel are given by
\begin{align}\label{eq:fd_f_c}
    \frac{ F_c{\K{q}} }{ 4\pi } &= \K{ 1 + (-1)^{l(\zeta_c) + s(\xi_c)} } \ibraket{c}{\xi_c}{\xi_n}{n} \nonumber \\
      &\quad \times \rint{\qp} X_{cn}{\K{q,\qp}} \tau_n{\K{q'}} F_n{\K{\qp}} \,, \\ 
    \frac{ F_n{\K{q}} }{ 4\pi } &= \ibraket{n}{\xi_n}{\xi_c}{c} \rint{\qp} X_{nc}{\K{q,q'}} \tau_{c}{\K{\qp}} F_c{\K{\qp}} \nonumber \\
      &\quad - \imel{n}{\xi_n}{\pmospin}{\xi_n}{n}  \nonumber \\
      &\quad\quad \times  \rint{\qp} X_{nn}{\K{q,q'}} \tau_n{\K{\qp}} F_n{\K{\qp}} \,, \label{eq:fd_f_n} 
\end{align}
whereby the functions \(F_i\) are related to the abstract components \(\ket{F_i}\) via \(F_i{\K{q}} \coloneqq \rint{\p} g_{l(\zeta_i)}{\K{p}} \ibraket{i}{p,q;\zeta_i}{F_i}{}\) with some orbital angular momentum quantum numbers \(\zeta_i\) and the regulator functions \(g_l\).
The functions \(\tau_i\) are closely related to $t$-matrix elements, while the ``kernel functions'' \(X_{ij}\) originate from the evaluation of free Green's functions between states differing in the spectator.
Thus they describe the one-particle exchange contribution in the three-body system.
Explicit expressions can be found in Ref. \cite{Hammer:2017tjm}.\footnote{
    Note that we use the definition
    \(X_{ij}{\K{q,q'}} \coloneqq \rint{\p} \rint{\pp} g_{l(\zeta_i)}{\K{\p}} g_{l(\zeta_j)}{\K{\pp}} \imel{i}{p,q; \zeta_i}{G_0}{\pp,\qp; \zeta_j}{j} \).
    In the case of sharp-cutoff regularization via the \(g_l\), these can be neglected at low momenta.
    If additionally the already mentioned interaction channels are \(s\)-wave, one can use
    the analytical expressions from Ref. \cite{Hammer:2017tjm}.
    The notation is slightly different, whereby the relation 
    \(X_{nc}{\K{q, q'}} = - m_n X_{00}^{n}{\K{q,q';B_3}}\) holds.
    Moreover, the relation \(X_{cn}{\K{q, q'}} = X_{nc}{\K{q', q}}\) can be employed.
    The function \(X_{nn}\) has a \(\pmospatial\) in front of the \(G_0\).
    Here the relation \(X_{nn}{\K{q, q'}} = - m_n X_{00}^{c}{\K{q,q';B_3}}\) can be used.
}

In order to renormalize the system to the physical binding energy of the three-body system, a three-body force is used.
It can be included by replacing \(X_{nn}{\K{q,q'}}\) by \(X_{nn}{\K{q,q'}} + h_3\) with \(h_3\) being some three-body force parameter, see, e.g., Ref. \cite{Hammer:2017tjm}.
Another quantity appearing in the Faddeev equations is the multiindex \(\xi_n\) specifying the spin state of the three-body system seen from the neutron as the spectator when the \(nc\) subsystem is in its interaction channel.
Analogously, \(\xi_c\) specifies the spin state seen from the core as the spectator when the \(nn\) subsystem is in its interacting channel.
In the case of a spinless core, these overlaps read
\begin{align}
    \ibraket{n}{\xi_n}{\xi_c}{c} &= -1 \,, \\
    \imel{n}{\xi_n}{\pmospin}{\xi_n}{n} &= -1 \,, 
\end{align}
and one obtains Faddeev equations equivalent to the ones from Ref. \cite{Canham:2008jd}.
We have explained this equivalence in detail in Ref.~\cite{Gobel:2022pvz}.

To calculate observables it can be efficient to use wave functions as an intermediate step.
This helps to modularize the computation.
As can be derived based on the abstract formalism (see, e.g., Refs. \cite{Ji:2014wta,Gobel:2019jba,Gobel:2022pvz}), the expression for the wave function component with an \(s\)-wave within the \(nn\) pair and an \(s\)-wave between this pair and the core is given by 
\begin{align}
    \Psi_c{\K{p,q}} &\coloneqq \ibraket{c}{p,q; \K{0,0}0,0; \xi_c}{ \Psi }{} \nonumber \\
    &= \psi_c{\K{p,q}} + \ibraket{c}{\xi_c}{\xi_n}{n} \int_{-1}^1 \dd{x} P_0{\K{x}} \nonumber \\
    &\quad \cross \psi_n{\K{\kcnp{\K{p,q,x}},\kcnq{\K{p,q,x}}}} \label{eq:Psi_c} \,,
\end{align}
where further details and explicit expressions for the functions
$\kcnp{\K{p,q,x}}$ and $\kcnq{\K{p,q,x}}$ can be found in Ref. \cite{Gobel:2019jba}.
Here the wave function of the full state in terms of Jacobi momenta relative to \(c\) as the spectator is given in terms of the component wave functions \( \psi_i{\K{p, q}} = G_0^{(i)}{\K{p, q}} \tau_i{\K{q}} F_i{\K{q}}\).
Despite having only \(s\)-wave interactions also the wave functions of the form \(\ibraket{c}{p,q; \K{l,l}0,0; \xi_c}{ \Psi }{}\), i.e., with non-zero subsystem orbital angular momenta, are non-zero (see, e.g., Ref. \cite{Gobel:2019jba}).
The numerical calculations, however, show that in the case of \(s\)-wave interactions these are strongly suppressed and negligible.

\subsection{$nn$ relative energy distribution}
\label{ssec:res_distribs}

The key object of our study is the ground-state \(nn\) relative-momentum distribution defined via the projection operator \(P_{p_{nn}}\) projecting onto a relative momentum \(p_{nn}\):
\begin{align}
    \rho\K{p_{nn}} &\coloneqq \expval{P_{p_{nn}}} \nonumber \\
    &= \int_{0}^{\Lambda} \dd{q} q^2 \Psi_c^2\K{p_{nn}, q} p_{nn}^2 \,. \label{eq:def_rho_p}
\end{align}
The \(nn\) relative-energy distribution follows from the momentum distribution via integration by substitution in the normalization integral:
\begin{equation}\label{eq:def_rho_E}
    \rho^{(E)}\K{E_{nn}} \coloneqq \sqrt{\frac{\mu_{nn}}{2E_{nn}}} \rho\K{\sqrt{2\mu_{nn}E_{nn}}} \,.
\end{equation}
However, measuring this distribution in the considered reaction is hardly possible, as it would require kinematics which suppresses not just all non-\(nn\) FSIs, but also \(nn\) FSI.
This requirement is not compatible with measuring the \(nn\) distribution at low relative momentum between the two neutrons. 
Therefore, we are mainly interested in the relative-energy distribution subsequent to \(nn\) FSI.
For this distribution the definitions in \cref{eq:def_rho_p,eq:def_rho_E} apply with the only 
difference that the ground-state wave function has to be replaced by the wave function after FSI, which is given by
\begin{align}
    &\Psi_c^{(\mathrm{wFSI})}{\K{p,q}} \nonumber \\
    &\quad \coloneqq \imel{c}{p,q; \zeta_c; \xi_c}{ \K{ \id + t_{nn}{\K{E_p}} G_0^{(nn)}{\K{E_p}} } }{\Psi}{} \nonumber \\
    &\quad = \Psi_c{\K{p,q}} + \frac{2}{\pi} g_0{\K{p}} \frac{1}{ a_{nn}^{-1} - r_{nn} p^2 /2 + \ci p} \nonumber \\
    &\quad\quad \times \rint{\pp} g_0{\K{\pp}} \K{p^2 - \pp[2] + \ci \epsilon}^{-1} \Psi_c{\K{p', q}} 
\end{align}
with \(\zeta_c \coloneqq (0,0)0,0 \).

\begin{figure}[htb]
  \centering
  \includegraphics[width=0.48\textwidth]{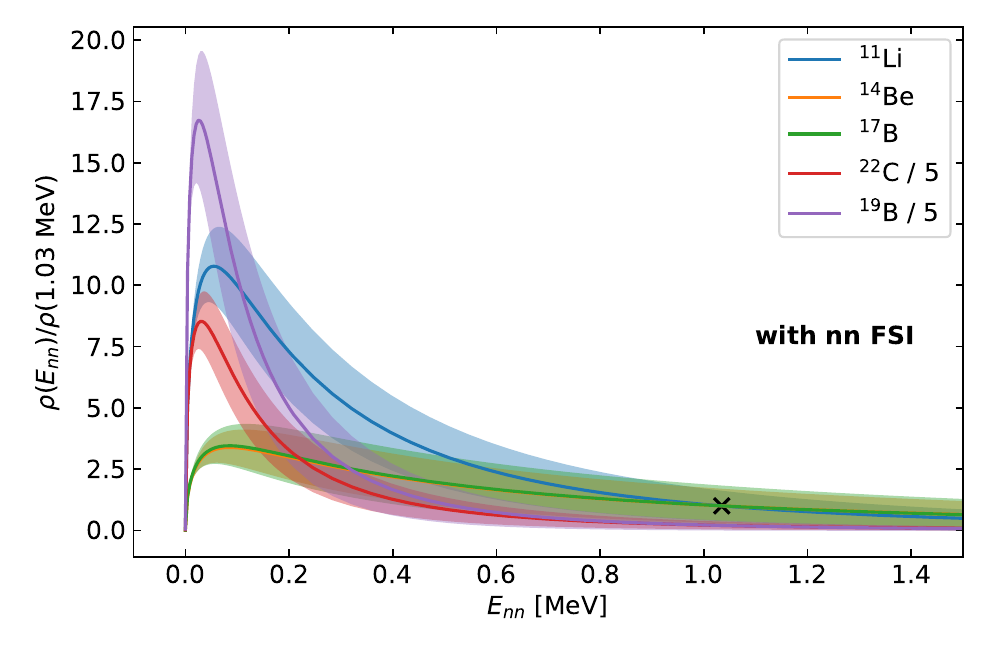}
  \caption{Plots of the \(nn\) relative-energy distributions of various halo nuclei with \(nn\) FSI included in the calculation. 
  The distributions are normalized to a certain value at a certain position
  indicated in the y-label and by the red cross.
  Leading-order EFT uncertainty bands are shown.
  The numerical uncertainties are smaller than 2\%.
  These uncertainties are estimated by comparing with a calculation having approximately twice as many mesh points
  and a 50\% larger cutoff.
  }
  \label{fig:E_nn_pd_cmp_rs_only_nnFSI}
\end{figure}

Details on how to evaluate this expression are given in, e.g., Ref. \cite{Gobel:2021pvw}.
The results of that evaluation are given in \Cref{fig:E_nn_pd_cmp_rs_only_nnFSI}.
The distributions are normalized such that they are unity at a certain energy around 1 MeV.
This normalization scheme is useful for comparison to experiment since the experimental distributions are measured for neutrons of low relative energy and the overall normalization is unknown. 

The plot shows that the results depend strongly on the considered nucleus;
there seems to be a great influence of \(S_{2n}\).
The lower it is, the larger the peak value of the distribution.
Only the results for \(^{14}\)Be and \(^{17}\)B are close to each other  (note that the orange curve is almost underneath the green curve), likely because of the similar two-neutron separation energies of the two nuclei.
Given the fact that the \(nc\) virtual state energies \(E_{nc}^*\) of these two nuclei differ by a factor greater than five, this indicates that, at least for these two systems, the scattering length of the \(nc\) interaction seems to be not so important.

%% file: tex/universality.tex

\section{Universality}
\label{sec:universality}

Next we discuss universality for the \(nn\) relative-energy distribution.
As we will see, there are different levels of universality that apply to this observable. 
Thus it is useful to start with a general discussion of the concept of universality to set the stage.

\textit{When is an observable universal?}
Loosely speaking, universality means that different systems display common features.
More specifically, universality can mean that some rescaled and/or differently parameterized version of the observable function is almost the same for the different systems.
In our case, the different systems are the different \(s\)-wave \(2n\) halo nuclei.

To make this discussion more concrete, we consider 
an observable \(\mathcal{O}(x)\), which is different for each of the \(N\) considered systems, i.e., it is \(\mathcal{O}(x; \vec{\theta})\), where the vector \(\vec{\theta}\) collects the characteristic parameters of the systems. 
This observable is universal if there is a rescaled version \( \mathcal{O}(x; \vec{\theta}) / f(\vec{\theta})\) which is almost independent of \(\vec{\theta}\), i.e.,
\begin{equation}\label{eq:univ_def_1}
  \mathcal{O}(x;\vec{\theta}) \approx f(\vec{\theta}) \widetilde{\mathcal{O}}(x) \,.
\end{equation}
One might, in addition, expect that \( f(\vec{\theta})\) varies only in a low-dimensional subspace of the full space of parameters \(\vec{\theta}\) that describe the system.
However, this is not essential to the observable being universal. 
Only in the case of \(f\) being additionally dependent on \(x\) would Eq.~\eqref{eq:univ_def_1} become pointless.
We note that the scaling variable may also be system-dependent, and so a slightly weaker form of universality is
\begin{equation}\label{eq:univ_def_2}
  \mathcal{O}(x;\vec{\theta}) \approx f(\vec{\theta}) \widetilde{\mathcal{O}}{\K{x / g(\vec{\theta})}} \, .
\end{equation}
For \(g(\vec{\theta})\) the same expectation as for \(f(\vec{\theta})\) holds: it should not be too complicated and should ideally 
depend only on a subset of the entries of \(\vec{\theta}\)---a subset that may be different than the subset present in \(f\).
Moreover, even if \(\widetilde{\mathcal{O}}\) would also depend on a subvector of \(\vec{\theta}\) denoted as \(\vec{\theta}_2\), i.e., \(\widetilde{\mathcal{O}}{\K{x / g(\vec{\theta});\vec{\theta}_2}}\), this would still be a variant of universality.

There are two equivalent ways to check universality based on a hypothesis for \(\widetilde{\mathcal{O}}\), \(f\), and \(g\).
The first is to compare the plots of \(f(\vec{\theta}) \widetilde{\mathcal{O}}{\K{x / g(\vec{\theta})}}\) as a function of \(x\) with the same plots of \( \mathcal{O}(x;\vec{\theta}) \) as a function of \(x\) for different systems.
This is basically testing the agreement between the universal prediction and the actual observable for all the systems.
Alternatively, one can plot
\begin{equation}
  \mathcal{O}(\tilde{x} g(\vec{\theta}); \vec{\theta}) / f(\vec{\theta}) \,,
\end{equation}
against \(\tilde{x} = x / g(\vec{\theta})\).
If there is universality as defined in \cref{eq:univ_def_2} amongst the \(N\) systems then the relation
\begin{equation}
  \mathcal{O}(\tilde{x} g(\vec{\theta}); \vec{\theta}) / f(\vec{\theta}) \approx \widetilde{\mathcal{O}}(\tilde{x})
\end{equation}
holds and the \(N\) curves for these different systems will lie approximately on one line.
That is a very illustrative way of testing universality: the curves obtained from different systems reduce to one, the universal prediction \(\widetilde{\mathcal{O}}{\K{\tilde{x}}}\).

\textit{The case of \(nn\) relative-energy distributions}
We are now in the position to apply these general considerations to the \(nn\) relative-energy distribution of $2n$ halo nuclei.
A key goal of this paper is to investigate if all the \(nn\) relative-energy distributions for \(s\)-wave \(2n\) halo nuclei can be derived by suitable (nucleus-dependent) rescalings of a single underlying function.

The first step in this direction is to find dimensionless distributions, where the independent variable, which we take to be the 
$nn$ pair's relative energy, \(E_{nn}\),  is measured in units of the characteristic energy of the system.
A natural choice for this energy is the two-neutron separation energy \(S_{2n}\), and indeed, we shall adopt  \(\tilde{x} \coloneqq E_{nn} / S_{2n}\).
The distribution \(\rho^{(\tilde{x})}{\K{\tilde{x}}}\) parameterized by \(\tilde{x}\) and abbreviated as \(\rho{\K{\tilde{x}}}\) or \(\rho{\K{E_{nn}/S_{2n}}}\) is given by
\begin{equation}
  \rho^{(\tilde{x})}{\K{\tilde{x}}} = \rho^{(E)}{\K{\tilde{x} S_{2n}}} \,,
\end{equation}
where $\rho^{(E)}$ is the relative energy distribution of the neutrons
(cf.~Eq.~\eqref{eq:def_rho_E}).
The normalization relation for this reparameterized distribution is
\begin{equation}
  \int_{0}^{\Lambda^2/\K{2\mu_{nn} S_{2n}}} \dd{\tilde{x}} \rho^{(\tilde{x})}{\K{\tilde{x}}} S_{2n} = 1 \,.
\end{equation}
This relation indicates that the size of the distribution \(\rho^{(\tilde{x})}{\K{\tilde{x}}}\) should scale as $1/S_{2n}$ for a given halo nucleus.
We will therefore take the function \(f(\vec{\theta})=S_{2n}\) here, while with the defintion of \(\tilde{x}\) we have already introduced \(g(\vec{\theta})=S_{2n}\).

We therefore take \(S_{2n} \rho^{(\tilde{x})}{\K{\tilde{x}}}\) as our universal observable \(\widetilde{\rho}\) and attempt to work out 
if any additional modifications to the alreday proposed \(f(\vec{\theta})\) and \(g(\vec{\theta})\) are required.
Moreover, we check if we can reduce the parameter vector \(\widetilde{\rho}\) depends on any further.

The possibility to describe  different \(2n\) halos in Halo EFT at leading order already manifests a certain kind of universality, since, e.g., the effective-range parameters of all the different \(nc\) interactions are only next-to-leading-order corrections and 
so become (approximately) irrelevant to the prediction. 
For the \(nn\) relative-energy distribution, LO Halo EFT implies:
\begin{align}\label{eq:rho_theta_3}
  &\widetilde{\rho}{\K{ E_{nn}/S_{2n}; \vec{\theta}_2 = (V_{nn}, V_{nc}, S_{2n}, A) }} \nonumber \\
  &\quad = \widetilde{\rho}{\K{ E_{nn}/S_{2n}; \vec{\theta}_3 = (\sqrt{2\mu S_{2n}} a_{nn}, \sqrt{2\mu S_{2n}} a_{nc}, A) }} 
\end{align}
with \(V_{nn}\) and \(V_{nc}\) containing all the \(nn\) and \(nc\) potential features the distribution might depend on.
Thus, in Eq. \eqref{eq:rho_theta_3} \(\vec{\theta}_3\) is an element of a three-dimensional subspace of the much larger parameter space containing \(\vec{\theta}_2\), i.e., \(\vec{\theta}_3 \subset \vec{\theta}_2\).
This ``Leading-order Halo EFT universality'' is a reduction-of-parameters statement.

Furthermore, systems with the same ratios of two-body and three-body scales, i.e., $\sqrt{2\mu S_{2n}} a_{nn}$ and $\sqrt{2\mu S_{2n}} a_{nc}$ will produce the same dimensionless distributions \(\widetilde{\rho}\). 
For systems to have the same values of these ratios would be somewhat coincidental were both numbers to be finite.
This situation is more likely to occur across several systems if either:
\begin{itemize}
\item $S_{2n}$ is large enough, in comparison to the energies of the \(nn\) and \(nc\) virtual states \(1/\K{2\mu a_{ij}^2}\), that $\sqrt{2\mu S_{2n}} a_{nn}$ and $\sqrt{2\mu S_{2n}} a_{nc}$ can both be taken to infinity. 
We denote this situation ``unitary universality''.

\item $a_{nc}$ is small enough, and $S_{2n}$ large enough, that we can work in the limit 
$\sqrt{2\mu S_{2n}} a_{nn} \rightarrow \infty$ and $\sqrt{2\mu S_{2n}} a_{nc} \rightarrow 0$. 
This is the limit in which \(nc\) interactions are a sub-leading effect.
In this case the \(nn\) halo is bound by a three-body force and is essentially a two-body (core-di-neutron) halo, since the 
\(nc\) pair is not dynamical.
This leads to a universal result for \(nn\) halo observables, as pointed out 
recently by Hongo and Son~\cite{Hongo:2022sdr}.
However, it is obviously a different limit to that considered under the previous bullet, and we denote 
it hereafter as ``nested two-body universality''. 
\end{itemize}

We will see below that there are a number of \(2n\) halos which exhibit this first case, ``unitary universality'', in their ground-state \(nn\) relative-energy distributions, i.e., 
\begin{align}
  &\widetilde{\rho}{\K{ E_{nn}/S_{2n}; \vec{\theta}_2 = (V_{nn}, V_{nc}, S_{2n}, A ) }} \nonumber \\ &\quad = \widetilde{\rho}{\K{ E_{nn}/S_{2n}; \vec{\theta}_4 = (A) }}
\end{align}
with \(\vec{\theta}_4 \subset \vec{\theta}_3 \subset \vec{\theta}_2\).
Moreover, we will find that in the unitary limit the $A$ dependence of this particular observable goes away too, leaving us 
with truly a one-parameter curve, that can be calculated in the double limit in which both \(nc\) and \(nn\) systems are at unitarity.
That curve approximately describes the ground-state \(nn\) relative-energy distribution of several \(2n\) halos quite accurately.

%% file: tex/main_part.tex

\section{Exposing the universal curve}
\label{sec:universalcurve}

\subsection{Universality of the ground-state relative-energy distribution}
\label{ssec:univ1}

\begin{figure}[htb]
  \centering
  \includegraphics[width=0.48\textwidth]{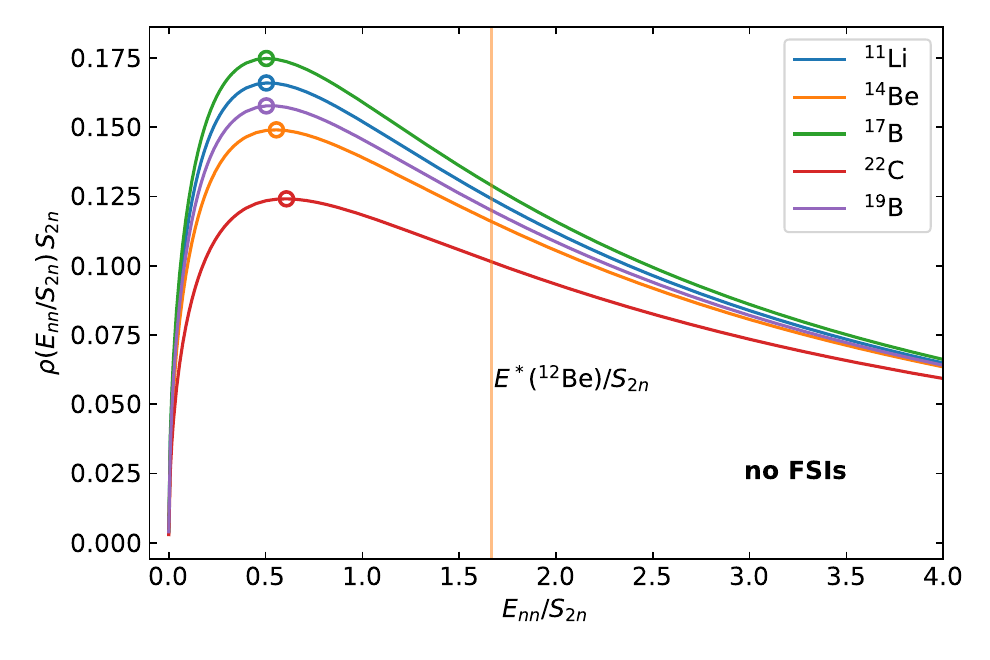}
  \caption{Plots of the ground-state \(nn\) relative-energy distributions times \(S_{2n}\) as a function
  of \(E_{nn} / S_{2n}\).
  The circles indicate the positions of the maxima.
  The vertical line indicates the breakdown scale of Halo EFT for \({}^{12}\)Be.
For all the other halo nuclei considered
  Halo EFT breaks down only beyond the region shown here.
  }
  \label{fig:rs_eps_nn_pd_cmp_no_ni_only_noFSI}
\end{figure}

At first glance, the results displayed in Fig.~\ref{fig:E_nn_pd_cmp_rs_only_nnFSI} do not support the existence of a universal result for the neutron energy distribution in two-neutron halo nuclei. 
We will show in this section that there is a common curve hidden within this result.
However, it is obscured because the \(nn\) FSI depends on a different set of variables than the ``unitary universality'' curve does. 
We therefore focus first on the distributions that are obtained 
before the inclusion of \(nn\) FSI.
This permits us to understand the ground-state dynamics better.
We first follow the argument from 
Sec.~\ref{sec:universality} and plot the distribution \(\widetilde{\rho}{\K{E_{nn}}} = S_{2n} \rho{\K{E_{nn}/S_{2n}}}\) in \cref{fig:rs_eps_nn_pd_cmp_no_ni_only_noFSI}.

We observe significantly more universality in these distributions without FSI: it is particularly noticeable that their maxima are now at approximately the same position.
But this only occurs if we plot versus \(E_{nn}/S_{2n}\).
The FSI is also universal, in the sense that the \(nn\) interaction that drives it is the same for all halo nuclei.
But there the relevant energy scale is set by the neutron-neutron scattering length, $1/(m_n a_{nn}^2)$. 
Our intermediate conclusion is that the universal description of these nuclei must be improved by taking into account the fact that the universal ground-state distribution and the universality due to the \(nn\) interaction introduce two different energy scales into the final, experimentally observed, distribution of the relative energy of the two neutrons. 

\begin{figure}[htb]
  \centering
  \includegraphics[width=0.48\textwidth]{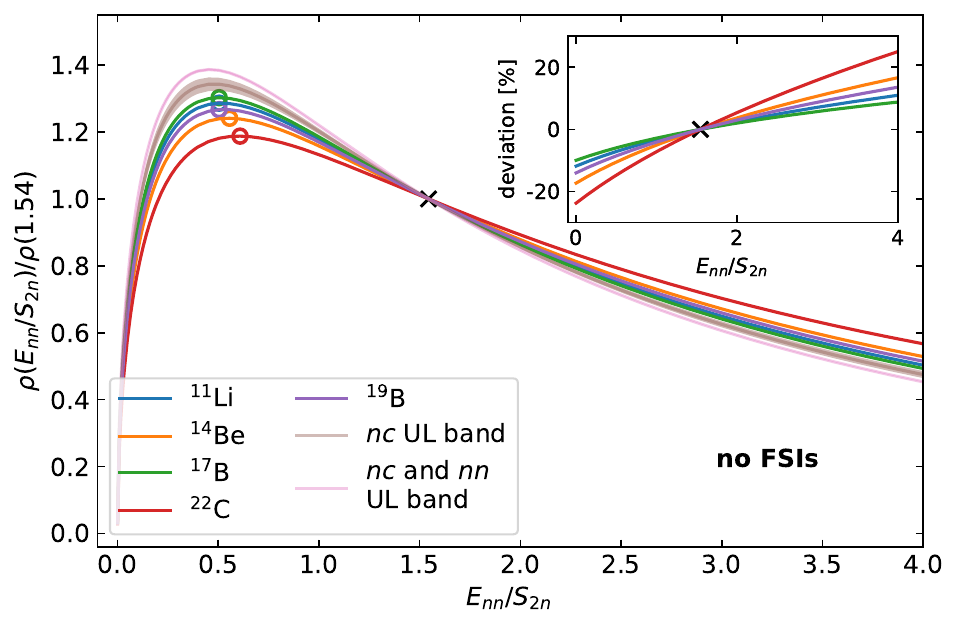}
  \caption{Ground-state \(nn\) relative-energy distributions from LO Halo EFT in comparison with two bands representing
  the calculations with the \(nc\) interaction in the unitary limit (brown) and the calculations with the \(nc\) and the \(nn\) 
  interactions in the unitary limits (pink).
  These calculations are represented by bands, as we varied \(A\) between 9 and 20.
  The normalization scheme where the distribution is normalized to a specific value at a specific point is used.
  The inset shows the deviations of the different curves from the doubly unitary limit in percent.
  }
  \label{fig:eps_nn_pd_nc_all_ul_cmp_band_only_noFSI}
\end{figure}

We begin by examining more closely the ground-state distribution.
First, we note that 
some of the \(nc\) virtual-state energies \(E_{nc}^*\) are not very accurately known.
But, in fact, this does not matter for our prediction.
If we push the \(nc\) interaction into the unitary limit, where the leading-order \(nc\) t-matrix simplifies from
\(
  \propto \K{\gamma_{nc}^{-1} + \ci p}^{-1}
\)
to
\(
  \propto \K{\ci p}^{-1}
\),
the ground-state energy distributions change by less than 25\%.
The results of such a calculation for the five different nuclei considered here are given by the brown band in \Cref{fig:eps_nn_pd_nc_all_ul_cmp_band_only_noFSI}.
(The band results from varying \(A\) in the region of interest for our work: between 9 and 20.)

We next investigate what happens if we do the same for the \(nn\) interaction, which also displays a large \(s\)-wave scattering length and is so not very far away from the unitary limit.
The result when both the \(nn\) and \(nc\) interaction are taken to the unitary limit is the pink curve in \Cref{fig:eps_nn_pd_nc_all_ul_cmp_band_only_noFSI}.

Note that we have {\it not} used the normalization scheme suggested by the arguments of the previous section. 
Instead, we followed the experimentally motivated prescription of Sec.~\ref{ssec:res_distribs}, and normalized to the same number at a particular value of \(E_{nn}/S_{2n}\)
 (\(E_{nn}=1.54 S_{2n}\)) in every curve.
The full LO results for the ground-state distribution, i.e., the results with interactions not taken to the unitary limit, are not far away from the bands.
In this range of \(E_{nn}/S_{2n}\) the deviations from the ``double unitary limit'' are at most 25\% and less than that for almost all nuclei over almost all the kinematic range.

This means that in addition to the LO EFT universality we can observe three further universalities:
\begin{enumerate}
  \item possibility to neglect the \(nc\) scattering length by going into the \(nc\) unitary limit,
  \item possibility to neglect the \(nn\) scattering length in the same way,
  \item possibility to neglect \(A\).
\end{enumerate}
These are all reduction-of-parameters universalities.
And they ultimately mean that to 25\% or better accuracy the quantity plotted in \Cref{fig:eps_nn_pd_nc_all_ul_cmp_band_only_noFSI} depends solely on the ratio
$E_{nn}/S_{2n}$ and not on any of the specific parameters---$a_{nc}$, $A$, or even $a_{nn}$---of the two-neutron halo. 
An important point regarding this universal distribution is that it describes the ratio \(\widetilde{\rho}(\tilde{x}=E_{nn}/S_{2n}) / \widetilde{\rho}(\tilde{x}_0)\), and not the probability density itself.
I.e., we are arguing that the shape of the distribution, with its overall size divided out, is a better candidate for a universal treatment.

\begin{figure}[htb]
  \centering
  \includegraphics[width=0.48\textwidth]{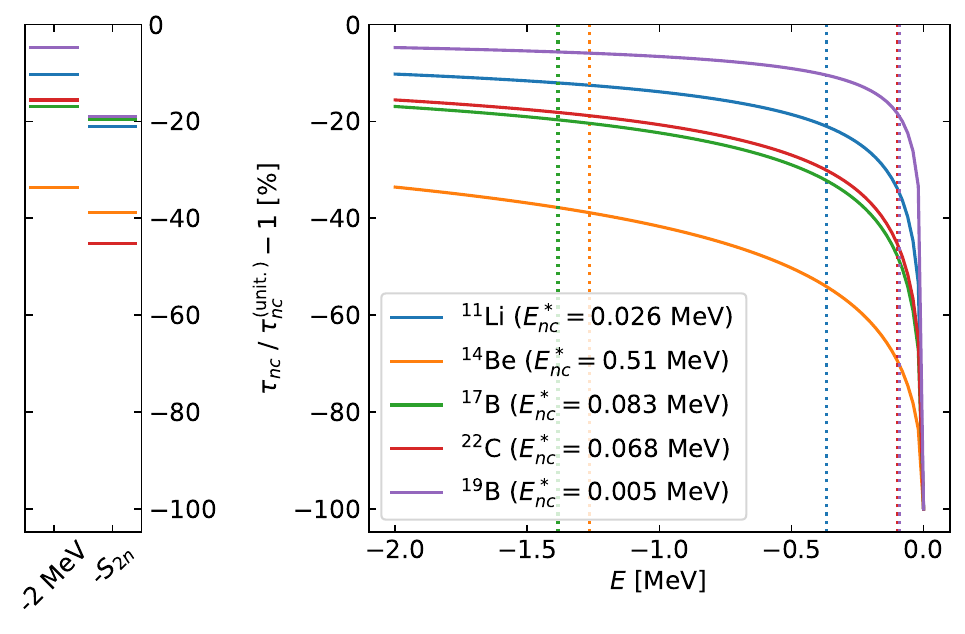}
  \caption{The deviation of the \(nc\) t-matrix from its unitary-limit variant.
  The vertical lines indicate minus the two-neutron separation energies, which represent the maximum
  energies at which the t-matrix is probed in the corresponding system.
  The small left panel indicates the deviations at these energies as well as at \(-2\) MeV.
  }
  \label{fig:tau_nc_dev_ul_E}
\end{figure}

The inset of \Cref{fig:eps_nn_pd_nc_all_ul_cmp_band_only_noFSI} shows the deviation of the neutron relative-energy distribution from the unitary-limit one.
To a significant extent the size of the deviations shown there can be understood in terms of the particular parameters that pertain to each Borromean \(2n\) halo.
The Faddeev equations \eqref{eq:fd_f_c} and \eqref{eq:fd_f_n} can be expressed in dimensionless variables if the input scattering lengths are recast as the dimensionless quantities \(\bar{a}_{nn}  \coloneqq \sqrt{2\mu S_{2n}} a_{nn}\) and \(\bar{a}_{nc}  \coloneqq \sqrt{2\mu S_{2n} a_{nc}}\). 
The larger these dimensionless scattering lengths are, the nearer each of the two t-matrices is to the corresponding unitary limit t-matrix.
The different ways the parameters \(S_{2n}\) and \(a_{nc}\) that go into \(\bar{a}_{nc}\) can bring the system nearer to the unitary limit can be understood from \cref{fig:tau_nc_dev_ul_E}.
It shows the deviation of the different halos' \(nc\) t-matrices in percent.
The t-matrices are plotted as a function of energy.
In the three-body calculation, they are evaluated at \(E=-S_{2n}-q^2/\K{2\mu_{n(nc)}}\).
Thereby, \(-S_{2n}\) determines the maximum energy at which the \(nc\) t-matrix can be probed.
Thus it can either be driven closer to the unitary limit by increasing the scattering length or restricting the probed region to more negative energies by increasing \(S_{2n}\).
In a dimensionless parameterization, these are two facets of the same process.
The figure illustrates both facets.

We note also that, since this is a three-body problem, the interplay of \(\bar{a}_{nc}\) and \(\bar{a}_{nn}\) is relevant.
The strength of one interaction can influence in which momentum region the other is probed.
A change of \(\bar{a}_{nn}\) could change the way the \(nc\) t-matrix is probed.
If it is probed more strongly at higher momenta the unitary limit approximation works better, without having changed either \(\bar{a}_{nc}\). 
The three-body force and the value of \(A\) also play a role in this regard.

Out of all the halos discussed here \(^{22}\)C is the one that has the largest deviation of its \(nc\) t-matrix from the unitary limit (UL) t-matrix at the two-neutron separation energy.
Moreover, also at higher energies, other nuclei display smaller deviations.
The larger deviations of the UL distribution of \(^{22}\)C from the LO result in \cref{fig:eps_nn_pd_nc_all_ul_cmp_band_only_noFSI}.
can already be understood in terms of the size of the dimensionless parameters \(\bar{a}_{nc}\) in that system.
Furthermore, we observe that the nucleus with the second largest unitary-limit deviation for the \(nc\)
interaction---\(^{14}\)Be---also has the second largest deviation for the ground-state relative-energy distribution.
However, while \(^{14}\)Be was quite prominent amongst the remaining nuclei in regard to its \(nc\) interaction deviation, the deviation of its distribution is not much larger than those of the remaining nuclei.
This highlights that also the three-body dynamics can have quite some influence.

The fact that the \(nn\) t-matrix can be moved into the unitary limit without changing the distributions
 very much could be understood in terms of a similar plot.
Only the three-body dynamics and the size of \(S_{2n}\) are relevant in that case, as the virtual-state energy of the \(nn\) system is the same for all nuclei. 
The deviations of the exact \(nn\) t-matrix from the corresponding unitary limit t-matrix at the two-neutron separation energies are roughly 23\% for \(^{14}\)Be and \(^{17}\)B, 36\% for \(^{11}\)Li, 52\% for \(^{22}\)C, and 53\% for \(^{19}\)B.
Therefore, due to having the smallest two-neutron separation energies, \(^{19}\)B and \(^{22}\)C are the ones with the highest deviations of the rescaled \(nn\) t-matrix in the region that is relevant for the bound-state calculation. 

Therefore, if we are willing to accept approximations of about 20\% (or a bit more for \(^{22}\)C) we can conclude that the ground-state relative-energy distributions of these halos display ``unitary universality'':
\begin{align}
  &\widetilde{\rho}{\K{ E_{nn}/S_{2n}; V_{nn}, V_{nc}, S_{2n}, A }} \nonumber \\
  &\quad = \widetilde{\rho}{\K{ E_{nn}/S_{2n}; A }} \nonumber \\
  &\quad = \widetilde{\rho}{\K{ E_{nn}/S_{2n}}}  \,.
\end{align}

\subsection{Universality with \(nn\) FSI included}
\label{sec:includingFSI}

Having discovered a remarkable amount of universality in the ground-state relative-energy distribution of two-neutron 
halos we now turn our attention to the impact that the \(nn\) FSI---and the corresponding energy scale \(1/(m_n a_{nn}^2)\)---has on 
that distribution.

First, we state the obvious: also taking the \(nn\) interaction for the FSI in the unitary limit is not a desirable approximation.
In the bound-state calculation the \(nn\) t-matrix is only ever evaluated at distances of at least \(S_{2n}\) away from the origin on the energy axis.
However, the FSI takes place on the positive real energy axis, and the \(nn\) t-matrix needs to be evaluated at energies of order \(1/(m_n a_{nn}^2)\).
Pushing the \(nn\) interaction to the unitary limit for the ground-state distribution is a good approximation for \(2n\) halos with \(S_{2n} \gg \frac{1}{m_n a_{nn}^2}\), but that approximation cannot be made in the low-energy region relevant for FSI.

The need to treat the two different components of the reaction calculation differently, due to the different energy regions that are probed in each case, means that 
we must find a way to build \(nn\) FSI on top of the ground-state result.
To take the FSI exactly into account one needs to apply the t-matrix on the wave function level.
But one can also employ the approximate technique of so-called FSI enhancement factors;
these can be directly applied to the distributions.
A detailed comparison between the exact calculation and FSI enhancement factors as well as an overview of different FSI enhancement factors can be found in Ref. \cite{Gobel:2021pvw} and its supplemental material
(see also Refs. \cite{Watson:1952ji,Slobodrian:1971an,boyd69}).

Our suggestion for a universal \(nn\) distribution including \(nn\) FSI is therefore
\begin{align}
  &\widetilde{\rho}^{(\mathrm{wFSI})}{\K{ E_{nn}/S_{2n}; V_{nn}, V_{nc}, S_{2n}, A }} \nonumber \\
  &\quad= \widetilde{\rho}{\K{ E_{nn}/S_{2n}}} G{\K{ a_{nn} \sqrt{2\mu E_{nn}}, r_{nn} \sqrt{2\mu E_{nn}} }}\,. \label{eq:univ_f_distrib}
\end{align}
Here \(G{\K{ a_{nn} \sqrt{2\mu E_{nn}}, r_{nn} \sqrt{2\mu E_{nn}} }}\) is an enhancement factor parameterized in terms of the 
dimensionless variables \(a_{nn} \sqrt{2\mu E_{nn}}\) and \(r_{nn} \sqrt{2\mu E_{nn}}\).
(The standard enhancement factors parameterized by the \(nn\) relative-momentum \(p_{nn}\), the \(nn\) scattering length \(a_{nn}\) and the \(nn\) effective range \(r_{nn}\) can, in fact, be rewritten exactly like this, as functions 
of the neutron-neutron scattering length and the effective range measured in units of the \(nn\) relative momentum.)
Popular enhancement factors are the \(G_1\) factor which can be found in Ref. \cite{Slobodrian:1971an}
and \(G_2\) from Ref. \cite{boyd69}.
It is worth noting that one cannot easily get rid of \(r_{nn}\) in these factors, as these two standard factors would diverge in the limit \(r_{nn} \to 0\).

\begin{figure}[htb]
  \centering
  \includegraphics[width=0.48\textwidth]{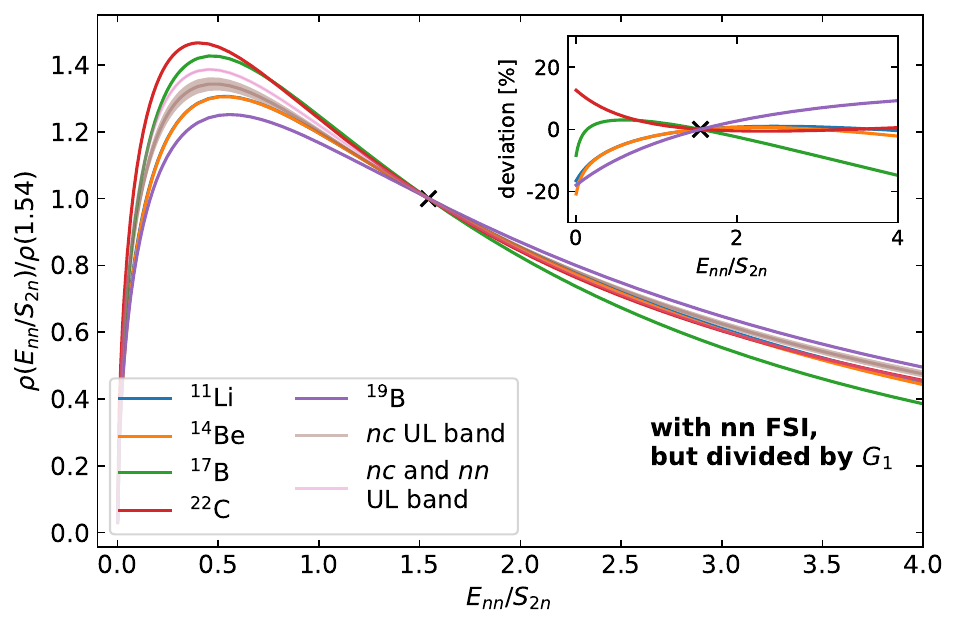}
  \caption{Full \(nn\) relative-energy distributions from LO Halo EFT divided by the FSI enhancement factor
  \(G_1\) in comparison with two bands representing the ground-state calculations with the \(nc\) interaction in the unitary limit (brown) and the calculations with the \(nc\) and the \(nn\) interactions in the unitary limits (pink).
  These calculations are represented by bands, as we varied \(A\) between 9 and 20.
  The normalization scheme where the distribution is normalized to a specific value at a specific point is used.
  The inset shows the deviations from the universal description.
  }
  \label{fig:eps_nn_pd_nc_all_ul_fsi_ef_cmp}
\end{figure}

Our universal description for the final distribution is not as universal as our result for the ground-state distribution.
It is different for each nucleus considered because we get an additional direct \(S_{2n}\) dependence from the enhancement factor if we stay in the \(E_{nn}/S_{2n}\) plot.
The enhancement factor's dimensionless variables \(a_{nn}\sqrt{2\mu E_{nn}}\) and \(r_{nn}\sqrt{2\mu E_{nn}}\) depend on \(E_{nn}\) but not \(E_{nn}/S_{2n}\) and thereby in that plot the factor is different for each nucleus.
If one wants to compare the exact calculations (or future experimental results) against the universal curve plus enhancement factor in the \(E_{nn}/S_{2n}\) plot, one would have to draw a new universal prediction for each nucleus.
But this plotting complexity can be eliminated by dividing the full LO Halo EFT calculation with FSI by the FSI enhancement factor and comparing the results obtained in this way with the ground-state universal curve.
That procedure is just an easier way of testing the statement from \cref{eq:univ_f_distrib}.
This comparison is shown in \cref{fig:eps_nn_pd_nc_all_ul_fsi_ef_cmp}.

It can be seen that this approach leads to good results.
In the case of all nuclei, the deviations are not significantly larger than 20\%.
Thereby, we have found a well-working universal description that is based on two different kinds of universality:
\begin{itemize}
  \item A reduction-of-parameter universality that eliminates the 
  dependence of the distribution on the nucleus under consideration: the EFT reduces the number of parameters
  which are then further reduced by being able to use unitary-limit amplitudes in the calculation of the bound-state distribution in our normalization scheme. The \(A\) dependence turns out to be negligible.
  \item A factorization universality: the FSI enhancement factor is the same for all nuclei. 
\end{itemize}

%% file: tex/conclusion.tex

\section{Conclusion}
\label{sec:conclusion}

We calculated the \(nn\) relative-energy distributions of various \(2n\) halo nuclei at leading order in Halo EFT. 
These can be observed in high-energy core-knockout reactions.
This was done at the ground-state level as well as after inclusion of \(nn\) FSI.
We found that good results for the ground-state distributions can be obtained by replacing the \(nc\) as well as the \(nn\) amplitude by the unitary limit amplitude, 
so turning the corresponding scattering lengths into superfluous parameters as far as the bound-state calculation is concerned.
Furthermore, with a suitable normalization of the distribution, the unitary-limit result is also rendered independent of the mass number of the nucleus. 

There is thus a universal ground-state \(nn\) relative-energy distribution for all \(s\)-wave Borromean \(2n\) halo nuclei.
We assessed the deviations of this universal result from the leading-order prediction of Halo EFT.
In the \(E_{nn}\) region from 0 to 4\(S_{2n}\) they are smaller than 20\% for the four halo nuclei \(^{11}\)Li, \(^{14}\)Be, \(^{17}\)B, and \(^{19}\)B, and slightly higher for 
 the case of \(^{22}\)C.
The neutron relative-energy distribution derived using ``unitary universality'' is as good an approximation as using leading-order Halo EFT.  This approximation is governed by the large values of the scattering lengths and the fact that, within the three-body (\(nnc\)) bound-state calculations, the corresponding t-matrices are only probed at negative energies with some distance to 0.

Georgoudis has also recently argued that the unitary limit is a useful way to 
understand the compound system of two neutrons and a heavy even-even nucleus~\cite{Georgoudis:2023dzp}.
However, the focus in that work is on energies and widths of states in the compound system. 

The unitary universality we found for neutron energy distributions across \(2n\) halos is not straightforward to observe experimentally, though, since it occurs at low relative momentum between the two neutrons.
In that kinematic region \(nn\) final-state interactions are strong.
And while the \(nn\) FSI is not affected by the mass of the core \(c\), or by the nature of the \(nc\) interaction, the universality of the \(nn\) FSI is not aligned with that of the ground-state distribution.
The former involves rescaling by the energy scale \(1/(m_n a_{nn}^2)\), while the 
latter involves rescaling by \(S_{2n}\). 

But the \(nn\) FSI can be taken into account by multiplying the universal \(nn\) relative-energy distribution by a final-state enhancement factor, i.e., writing:
\begin{align}
    & \rho^{(\mathrm{wFSI})}(E_{nn}) \nonumber\\
    & \quad = \widetilde{\rho}{\K{ E_{nn}/S_{2n}}} G{\K{ a_{nn} \sqrt{2\mu E_{nn}}, r_{nn} \sqrt{2\mu E_{nn}} }}\,. 
\end{align}
We assessed the accuracy of this approximation for the final distribution by comparing it to the leading-order EFT calculations with \(nn\) FSI taken into account using the \(nn\) t-matrix and found differences of less than 20\% almost everywhere.

In the future, it would be interesting to extend the Halo EFT calculation to next-to-leading order and 
see how large deviations from the universal results are.
This could be complemented by tests in higher-resolution three-body models and {\it ab initio} calculations based on 
chiral effective field theory.

Moreover, it would be very interesting to measure these distributions experimentally.
This would provide another experimental test case for leading-order Halo EFT, but, more than that, would 
permit an empirical evaluation of the claim that there is a universal \(nn\) relative-energy distribution that occurs in all these \(2n\) halos. 
Such measurements of the \(nn\) relative-energy distributions following a high-energy core knockout are, in principle, feasible at rare isotope beam factories.
Indeed, one was already done using the missing-mass technique for the calibration reaction \(^6\)He\((p,p'\alpha)nn\) of the recent RIKEN experiment regarding the tetraneutron \cite{Duer:2022ehf}.
A kinematically complete measurement of the same reaction will be performed at RIKEN for the purpose of measuring the \(nn\) scattering length \cite{Gobel:2021pvw,nn_scat_len_ribf_prop2018}.